# Effect of alloying on the microstructure, phase stability, hardness and partitioning behavior of a new dual-superlattice nickel-based superalloy

C. Rodenkirchen[1], A. K. Ackerman[1], P. M. Mignanelli[2], A. Cliff[1], G. J. Wise[4], J. O. Douglas[1], P. A. J. Bagot[3], M. P. Moody[3], M. Appleton[2], M.P Ryan[1], M. C. Hardy[2], S. Pedrazzini[1]*, H. J. Stone[4]

[1] Department of Materials, Imperial College London, Royal School of Mines, Exhibition Road, London, SW7 2AZ, UK
[2] Rolls-Royce plc, PO box 31, Derby, DE24 8BJ, UK
[3] Department of Materials, University of Oxford, Parks Road, Oxford, OX1 3PH, UK
[4] Department of Materials Science and Metallurgy, University of Cambridge, 27 Charles Babbage Road, Cambridge, CB3 0FS, UK

*corresponding author, email: s.pedrazzini@imperial.ac.uk

## Abstract

A novel γ-γ′-γ″ dual-superlattice superalloy, with promising mechanical properties up to elevated temperatures was recently reported. The present work employs state of the art chemical and spatial characterization techniques to study the effect systematic additions of Mo, W and Fe and variations in Nb and Al contents have on the phase fraction, thermal stability, elemental partitioning and mechanical properties. Alloys were produced through arc melting followed by heat treatment. Multi-scale characterization techniques and hardness testing were employed to characterize their microstructure, thermal stability and mechanical properties. Alterations in such properties or in elemental partitioning behavior were then explained through thermodynamic modelling.

A modest addition of 1.8 at.% Mo had a strong effect on the microstructure and thermal stability: it minimized microstructural coarsening during heat treatments while not significantly decreasing the γ′ solvus temperature. A reduction of Nb by 0.6 at.%, strongly reduced the γ″ volume fraction, without affecting the γ′ volume fraction. The reduced precipitate fraction led to a significant reduction in alloy hardness. Fe, added to achieve better processability and reduced material cost, decreased the γ′ solvus temperature and caused rapid microstructural coarsening during heat treatments, without affecting alloy hardness. A reduction of Al by 0.4 at.%, reduced the γ′ volume fraction and the γ′ solvus temperature, also without affecting alloy hardness. The addition of 0.9 at.% W decreased the γ′ solvus temperature but increased both precipitate volume fractions. These data will be invaluable to optimize current alloy design and to inform future alloy design efforts.

## 1. Introduction

There is a global drive to improve engine efficiency and reduce emissions in civil aviation. This is motivating the development of new nickel-based superalloys capable of withstanding the higher operating temperatures and increased rotational speeds demanded in next generation of gas-turbine engine designs [1]. Currently, Inconel® 718 (Ni-20.5Cr-18.85Fe-0.33Co-1.84Mo-1.18Al-3.3Nb-1.23Ti-0.34C-0.02B at.%) [2] remains the most widely used alloy for intermediate temperature applications due to its favorable combination of processability and mechanical strength up to 650 °C. Indeed, it constitutes approximately one third of the weight of some engines [3]. Inconel® 718 is strengthened by precipitation of two ordered intermetallic phases, ~3% γ′ (based on $Ni_3Al$ with the $L1_2$ structure) and ~20% γ″ (based on $Ni_3Nb$ with the $D0_{22}$ structure) embedded coherently within the A1 γ nickel matrix [4]. The γ″ phase confers additional strengthening over the γ′ phase alone (on a normalized volume fraction basis) due to the greater coherency strain resulting from the lattice misfit between the tetragonal *c*-axis of the γ″ and the γ matrix. However, the

γ″ phase is also metastable, and rapidly transforms to the thermodynamically stable orthorhombic D0$_a$ δ phase at temperatures above 650°C, limiting the upper operational temperature [5][6].

Alloys with superior temperature capability and strength to Inconel® 718 have been developed, although such alloys typically rely on higher refractory metal contents and precipitate volume fractions. Consequently, these alloys possess narrow processing windows, exhibit rapid work hardening, are not readily weldable and are susceptible to hot shortness [6–10]. Whilst such alloys may be processed using powder metallurgical routes, this significantly increases alloy costs compared to cast and wrought products. As a result, there remains significant interest in developing affordable cast and wrought alloys with superior properties to Inconel 718®, whilst simultaneously retaining the desirable processability characteristics.

Efforts to enhance the temperature capabilities of 718-type alloys have investigated compositional modifications as well as carefully defined heat treatment schedules to modify the precipitate morphology. Such efforts have been successful, but have only led to relatively modest improvements in thermal stability [4][12]. More recently, studies of alloys based on Ni-Cr-Al-Nb have identified the possibility of alloy microstructures containing appreciable volume fractions of both γ′ and γ″ [6][13]. These alloys were reported to have yield strengths >100 MPa higher than many other Ni-based superalloys. Subsequent work has demonstrated that these alloys retain microstructural stability up to 750˚C [14] and that further improvements to their properties may be achieved through judicious alloying [8]. However, if these alloys are to be effectively optimized for high temperature service, a detailed understanding must be established of how alloying additions affect the relative stability, morphology and properties of the constituent phases.

The original studies of dual superlattice superalloys reported that an alloy with composition of Ni-15Cr-4Al-6Nb (at.%) contained ~25 γ′ and ~20% γ″. Whilst this alloy demonstrated attractive properties, three alloy development opportunities may be identified based on established superalloy metallurgy. First, the phase fractions are higher than many other superalloys derived from Inconel® 718. This may adversely affect the processability of the alloy. It is therefore appropriate to consider alloys with lower Al and Nb contents to determine the extent to which the beneficial properties of a dual-superlattice microstructure may be retained with lower precipitate fractions. Second, the quaternary composition affords comparatively limited solid solution strengthening of the γ matrix. As such, further benefits may be derived from considering additions of elements such as Mo and W to provide effective strengthening [15]. Third, the elemental cost of the dual superlattice alloy is higher than Inconel® 718. It is therefore appropriate to consider the extent to which Fe substitutions may be made as this will significantly reduce alloy cost and ease recycling.

Previous studies have shown that Mo and W additions partition strongly to the γ matrix, where they provide potent solid solution strengthening on a per atomic percent basis [16,17]. W preferentially sits at Al sites in γ′, but if Ta is present in the alloy system, W will instead partition to the γ matrix [18]. However, these elements are also known to promote the formation of topologically close-packed phases (TCPs) such as σ and μ phase [19,20][21] which are considered deleterious and lead to reduced creep lives. The concentrations of these elements must therefore be carefully controlled to ensure microstructural stability.

In sufficient concentrations, Nb promotes the formation of Ni$_3$Nb γ″ precipitates, as well as contributing to the γ′ precipitates where it substitutes for aluminum. As it does not partition strongly to the γ′, Nb additions usually have a limited effect on the γ-γ′ lattice misfit [22]. Nb also increases the anti-phase boundary (APB) energy of the γ′, as well as acting as a solid solution strengthener in both the γ matrix and γ′ precipitates [23],[22]. The addition of Nb has also been reported to reduce precipitate coarsening rates, which has been attributed to its sluggish diffusion kinetics in Ni [24].

Fe is an important constituent of Inconel® 718 and related alloys. It significantly reduces raw material costs, whilst simultaneously improving processability and recyclability. However, if the content is too high, deleterious Laves phases (Ni,Fe,Cr)$_2$(Nb,Mo,Ti) can be formed, which compromise the mechanical properties. Fe has also been reported to aid in the formation of γ″ precipitates, providing the necessary electron:atom ratio for stable D0$_{22}$ precipitation [25].

The present study involves a combination of microstructural and chemical characterization experiments and thermodynamic modelling, used to understand the properties of the resulting alloys. Six systematic compositional variations of the dual superlattice alloy were produced to study the effect of Mo, Nb, W and Fe on the phase stability, solvus temperature, and elemental phase partitioning. The results obtained give insight into the extent to which each of these additions may be accommodated without compromising the alloy microstructure and may therefore help inform future development of these alloys.

## 2. Experimental Methods

Six polycrystalline nickel-based superalloys were produced through vacuum arc melting from raw elements of >99.9 wt.% purity, the alloy designations and nominal compositions of which are shown in Table I. The base alloy was selected to contain the same Cr, Al and Nb contents of the original dual superlattice superalloy investigated in [13]. In addition, small concentrations of C, B and Zr were included to provide grain boundary strengthening, in line with other polycrystalline Ni-based superalloys. The variant *+1.8Mo* included 1.8 at.% Mo, a composition similar to that in Inconel® 718. In the alloy *–0.6Nb* the Nb composition was reduced to 5.4 at.%, a reduction of 10% of the overall Nb content. The third variation, *+9Fe*, included 9 at.% Fe, similar to the composition in ATI 718Plus®. In the *–0.4Al* alloy, the Al content was reduced to 3.6%, a reduction of 10% of the overall Al content. Finally, *+0.9W* included 0.9 at.% W, a composition corresponding to ~3 wt.%. In this alloy, the Fe was removed in order to avoid complications arising between these two additions. Compositions were measured by doing five large area EDX maps on the alloys, the averaged data can be found in the appendix Table AI.

*Table I: Alloy designations and nominal chemical compositions in atomic % of the six nickel-based superalloys studied in the present work* [26].

| Name | Ni | Cr | Fe | Mo | W | Al | Nb | C | B | Zr |
|---|---|---|---|---|---|---|---|---|---|---|
| Base | bal. | 15 | - | - | - | 4 | 6 | 0.15 | 0.16 | 0.04 |
| +1.8Mo | bal. | 15 | - | 1.8 | - | 4 | 6 | 0.15 | 0.16 | 0.04 |
| -0.6Nb | bal. | 15 | - | 1.8 | - | 4 | 5.4 | 0.15 | 0.16 | 0.04 |
| +9Fe | bal. | 15 | 9 | 1.8 | - | 4 | 5.4 | 0.15 | 0.16 | 0.04 |
| -0.4Al | bal. | 15 | 9 | 1.8 | - | 3.6 | 5.4 | 0.15 | 0.16 | 0.04 |
| +0.9W | bal. | 15 | - | 1.8 | 0.9 | 3.6 | 5.4 | 0.15 | 0.16 | 0.04 |

All six alloys were prepared for heat treatment by encapsulation in quartz ampoules under vacuum. Solution heat treatments to improve microstructural homogeneity and remove solidification induced microsegregation were performed at 1200°C for 24 hours. Subsequent heat treatments were performed at 800 °C for 10 and 100 hours to induce precipitation and coarsening of the dual-superlattice $\gamma'$ and $\gamma''$ microstructures.

The effect of the different alloying additions on the $\gamma'$ solvus temperature was determined through differential scanning calorimetry (DSC). DSC thermograms were collected from samples after 100 h heat treatment at 800 °C, measuring 5 mm in diameter, and 1 mm thick, using a Netzsch 404 high-temperature calorimeter. The heat flux was measured during heating at 10°C min$^{-1}$ between 600 and 1400°C. The tests were performed under flowing Ar to minimize sample oxidation. The resulting data were analyzed according to the NIST guidelines [27] to determine the $\gamma'$ solvus temperatures.

Samples were prepared for metallographic examination by sectioning with a Buehler IsoMet saw, then mounted in conductive phenolic resin, ground using SiC papers, polished with progressively finer diamond grades, before a final chemical polish using a 0.04 μm oxide particle solution (colloidal silica). Samples were then electro-etched using a 10 vol.% phosphoric acid solution in distilled water at an applied voltage between 3-5 V. Scanning electron micrographs were obtained using an FEI Nova NanoSEM 450 and a Zeiss Auriga SEM, both operated at 5 kV.

X-Ray diffraction (XRD) was performed using a Bruker D8 Advance Gen.9 θ-2θ diffractometer fitted with a LynxEye EX position sensitive detector (PSD). Diffraction data were acquired using Cu-Kα radiation, generated using a 40 kV accelerating voltage, a 40 mA current and a 2θ collection range of 20-100°, with an increment of 0.02° and a dwell time of 7 s. Diffractograms were analyzed using CrystalDiffract software by comparing them to reference spectra available through the Inorganic Chemical Structures Database (© FIZ Karlsruhe).

Hardness measurements were carried out on a Wilson® VH1202 micro hardness tester using a load of 2 kg and dwell time of 15 s. For each sample, 21 measurements were taken from various points spread over the full sample area and the mean values and standard deviations were calculated.

Samples were prepared for Atom Probe Tomography (APT) through cutting and electropolishing of matchsticks, performed using first a solution of 15 vol.% perchloric acid in acetic acid and a 16-20 V DC voltage, then sharpened using a 2 vol.% perchloric acid in butoxyethanol at 10 V DC voltage [28]. Samples were analyzed in a Cameca LEAP 5000 XR atom probe, at 50 K, with a pulse frequency of 200 kHz, a detection rate of 1.5 %, and a pulse fraction of 0.2 for specimens run in voltage mode. Reconstructions were performed using the Cameca Integrated Visualization and Analysis Software (IVAS) 3.6.12 and AP Suite 6, using mainly standard voltage curve reconstruction. Constant shank angle reconstructions were used and compared to the voltage curves. These were only employed in the event the voltage curve reconstruction visibly distorted precipitates due to artifacts (e.g., to mitigate the effect of potential small sample fractures during analysis). Iso-concentration surfaces were created at 7 at.% Al and 13-15 at.% Nb to identify the different phases present in the alloys. Where a peak overlap was observed (e.g. Cr-Al-Fe at 27 Da, Ni-Nb at 31 Da), it was labelled as the phase with only one isotope, in this case Al and Nb. The reason for this is while Cr, Fe and Ni have other labelled primary isotope peaks, Al and Nb do not and therefore would not have been included in the reconstruction. Peak decomposition algorithms were then used to obtain phase composition values from isolated single-phase spectra, thereby reducing errors arising from peak overlaps.

Predictions of the phase fractions and phase compositions were completed using the Thermo-Calc 2019b software package, with the TTNi8 v8.3 Ni-alloy databases. All compositional predictions were obtained at an isothermal temperature of 800°C [29].

## 3. Results

DSC thermograms from all the alloys are shown in Figure 1. Both precipitate phases γ′ and γ″ strengthen the alloy, therefore knowledge of their solvus is important to understand the limits in operational temperature of the alloys. In this work, we have focused on understanding the effects on γ′ phase stability, which is particularly useful for high temperature applications. The stability of the γ″ phase will form part of a future work. The measured γ′ solvus temperature is summarized and compared to thermodynamic predictions in Figure 2 (values also given in appendix Table AII). The addition of 1.8 at.% Mo did not significantly alter the γ′ solvus temperature as it led to a reduction of only 1 °C from 1010 °C to 1009 °C, which lies within the experimental error of the DSC. The reduction of Nb content to 5.4 at.% reduced it by further 4 °C to 1005 °C and the addition of 9 at.% Fe resulted in a decrease in γ′ solvus temperature of 17 °C from 1005 °C to 988 °C. A further 18 °C reduction to 970 °C was observed for the -0.4Al alloy, which has an Al content of 3.6 at.%, 10% lower than the 4 at.% of the previous alloys. Al and Nb hence increase the γ′ solvus temperature whereas Mo and Fe reduce it. Lastly, the +0.9W alloy exhibited a γ′ solvus temperature of 983 °C. It is not as straightforward to determine the effect of W on the solvus temperature as it was for the other alloying additions. This is because the +0.9W alloy differs in composition of at least two elements compared with any of the other investigated alloys. The +0.9W alloy is best compared to either the -0.4Al alloy with a lower γ′ solvus temperature of 970 °C or to the -0.6Nb alloy with a higher γ′ solvus temperature of 1005 °C.

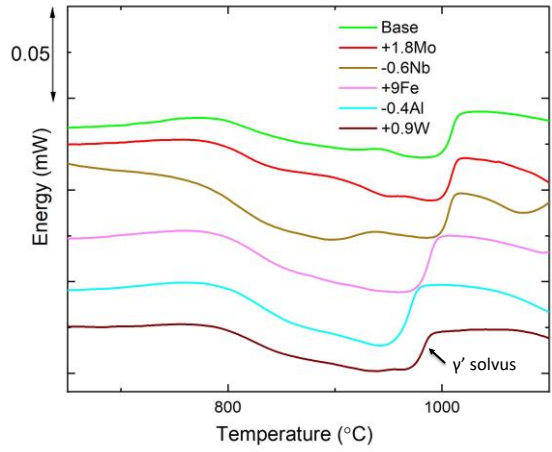

*Figure 1: DSC thermograms of the alloys, showing how individual alloying additions influence the γ' solvus temperature.*

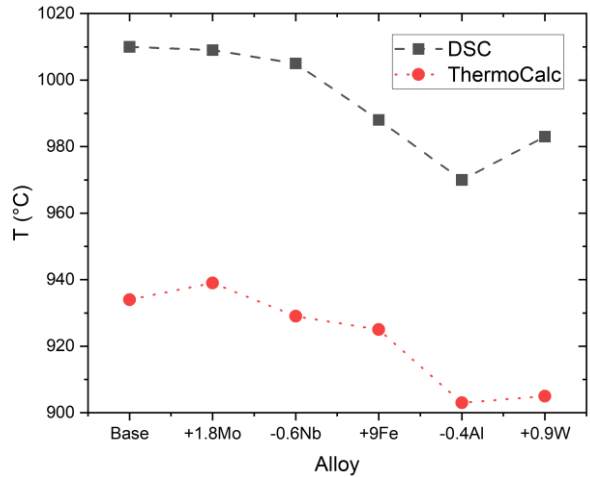

*Figure 2: Comparison of the γ' solvus temperatures of the alloys as derived from the DSC thermograms and as predicted by Thermo-Calc.*

First, a comparison between the -0.4Al alloy and the +0.9W alloy is undertaken. The difference between these two alloys is a removal of the 9 at.% Fe together with the addition of 0.9 at.% W. As previously stated, the addition of Fe resulted in a reduction of the solvus temperature. Hence, it is not surprising that the γ′ solvus temperature increases as the Fe is removed in the +0.9W alloy. However, the total difference in solvus temperature from the -0.4Al to the +0.9W alloy amounts to only 13 °C, which is substantially less than what was observed for the addition of 9 at.% Fe to the base alloy (17 °C). Hence, from the comparison of the -0.4Al alloy and the +0.9W alloy, it can be suggested that W reduces the γ′ solvus temperature.

The difference between the -0.6Nb alloy and the +0.9W alloy is a decrease of the Al content from 4 to 3.6 at.% alongside the addition of 0.9 at.% W. A comparison of these alloys reveals a strong reduction in γ′ solvus temperature of 22 °C. Comparing the base alloy and the –0.4Al alloy, it was found that the reduction of Al led to a reduction of the solvus temperature by 18 °C. Again, the conclusion can be drawn that W led to a further reduction of the γ′ solvus temperature. For both comparisons, the excess reduction attributed to the addition of 0.9 at.% W amounts to 4 °C.

The results collectively indicate that the γ′ solvus temperature is reduced by Mo, Fe, and W additions, among which W has the relatively strongest effect. However, more pronounced effects are observed for the alloying elements Nb and Al, which both increase the γ′ solvus temperature, with Al being by far the most potent element of all investigated elements. The potencies of these elements are summarized in Table II. The thermodynamic predictions from Thermo-Calc capture the overall trend of γ′ solvus temperature, particularly the correlation with Al content. However, the predicted effects of Mo and W do not appear to correlate with the experimental observations. Across the entire series of alloys, it is noted that there is an offset of approximately 70°C between the Thermo-Calc predictions and the experimental values.

*Table II: Summary of the potency of the alloying additions to affect the γ′ solvus temperatures.*

| Alloying element | γ′ solvus temperature potency |
|---|---|
| Mo | -0.6 °C/at.% |
| Nb | +6.7 °C/at.% |
| Fe | -1.9 °C/at.% |
| Al | +45 °C/at.% |
| W | -4.4°C/at.% |

A homogenizing heat treatment was chosen for all alloys at 1200°C in order to dissolve all precipitates, including γ′ and γ″, that may have formed during fabrication. This temperature was selected based on the DSC data in Figure 1, being above all solvus events and below the solidus for each alloy. Then, samples were heat treated at 800 °C to induce controlled levels of precipitation. Figure 3 shows the X-ray diffractograms produced after 100 hours at 800 °C. The base alloy was previously shown to precipitate primarily a combination of γ′, γ″ and δ phases, depending on the heat treatment [14]. After up to 1000 hours at 750°C, only γ, γ′ and γ″ were observed by synchrotron X-ray diffraction while at 800°C δ precipitation began to occur when the heat treatment exceeded 100 hours in duration [14].

In the present case, X-ray diffractograms, shown in Figure 3, confirmed that all alloys contained γ, γ′ and γ″ phases. The presence of the δ phase and niobium nitrite (NbN) was also confirmed in most alloys. The reduction of Nb may have reduced δ and NbN precipitation, whereas the reduction in Al and the addition of W promoted precipitation of these phases.

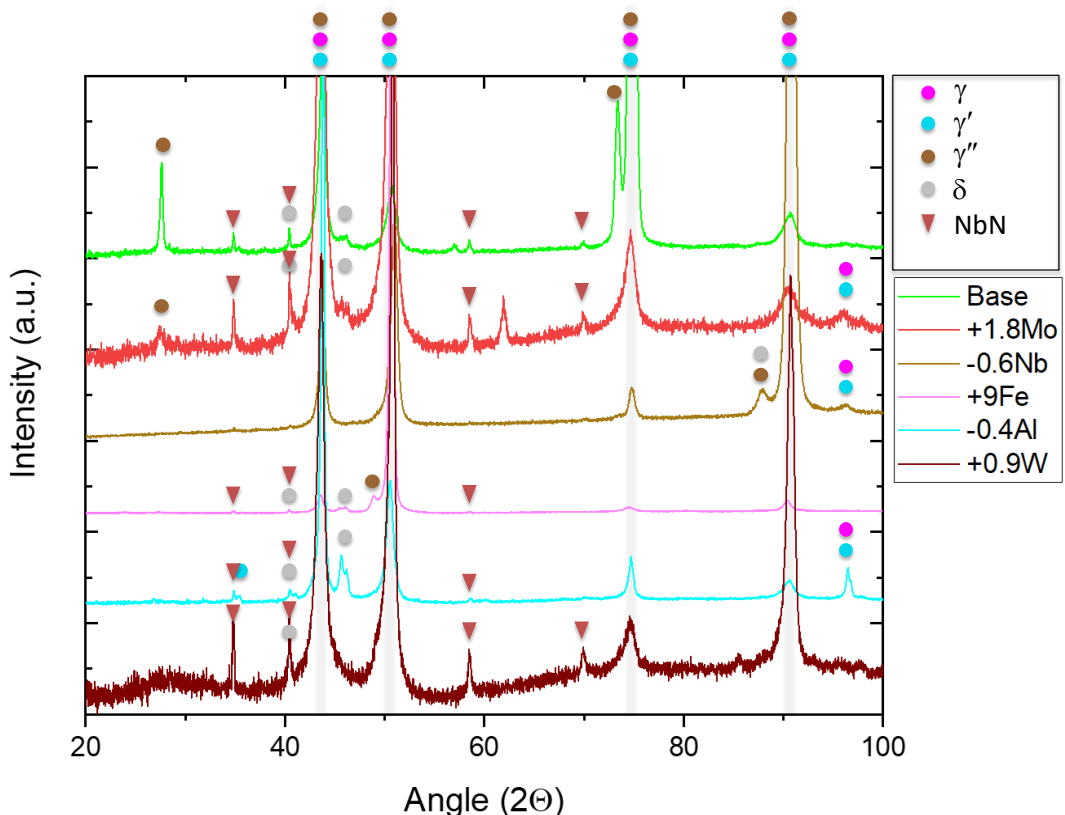

*Figure 3: X-ray diffractograms of the alloys after heat treatment for 100 hours at 800 °C.*

The measured hardness of the alloys is given in Figure 4. The most prominent effect is found for the reduction of Nb by 0.6 at.%, which led to a reduction in the hardness of over 40 HV. In contrast, the addition of 9 at.% Fe as well as the reduction of Al by 0.4 at.% did not significantly affect the alloy hardness. Lastly, the addition of the solid solution strengtheners Mo and W seem to have had a positive effect on the alloy hardness, although the relatively large standard deviations reduce the clarity of this observation.

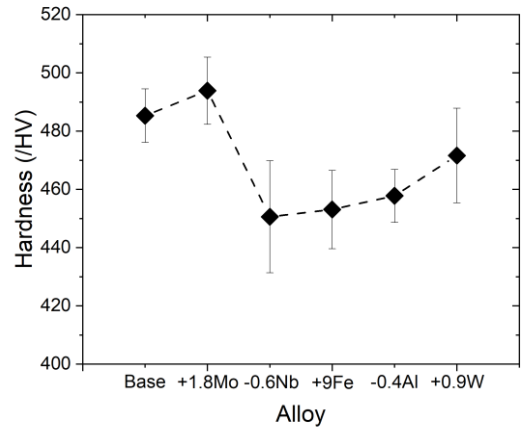

*Figure 4: Hardness data for all alloys after heat treatment for 100 h at 800 °C, averaged over 21 measurements.*

SEM micrographs of the alloys following heat treatment at 800 °C for 10 hours and for 100 hours are presented in Figure 5(a) and (b) respectively. The micrographs were taken from within the grains, away from grain boundaries. The δ phase is not seen in the presented micrographs, as it was found only in small quantities, localized at grain boundaries. Niobium nitrites were also found only in small quantities. The formation of the δ phase with respect to alloying alterations is beyond the scope of the present study and will be part of a future work.

As shown in Figure 5, the 10-hour heat treatment yielded extremely fine nano-structured microstructures. Heat treating for 100 hours coarsened the precipitates and allowed the effect of individual alloying additions on the precipitate morphology and fraction to become clearer. The presented micrographs were taken from within the grains, therefore no δ phase is shown, as this phase only precipitated at grain boundaries. The micrograph of the base alloy shows the precipitation of two distinct precipitate phases: cuboidal γ′ precipitates with rounded corners and thinner, lenticular γ″ precipitates. Both precipitates can also be observed in all the additional alloys, though changes in their morphology and distribution are found for each compositional variation.

The addition of 1.8 at.% Mo had a refining effect on the microstructure and the 0.6 at.% reduction of Nb seems to have led to a reduction in the precipitate phase fraction, both of which can be observed after 10 hours and even more so after 100 hours of heat treatment. Alloying with 9 at.% Fe led to substantial coarsening of the microstructure and the formerly cuboidal precipitates are now more irregularly shaped. The clear morphology difference between the γ′ and the γ″ precipitates is reduced as these phases seem to have grown into each other instead of being separated from one another by the matrix phase. With the reduction of the Al content in the -0.4Al alloy, the precipitate phase fraction further decreased and the precipitate phases were still clustered together rather than separated by the matrix. Finally, the removal of Fe and the addition of W in the +0.9W alloy restored the more refined microstructure similar to the +1.8Mo alloy, with more distinct precipitate phases than in the Fe containing alloys. The exact phase fractions of γ′ and γ″ of the alloys cannot be calculated from the SEM micrographs due to the similarity in contrast between the precipitates.

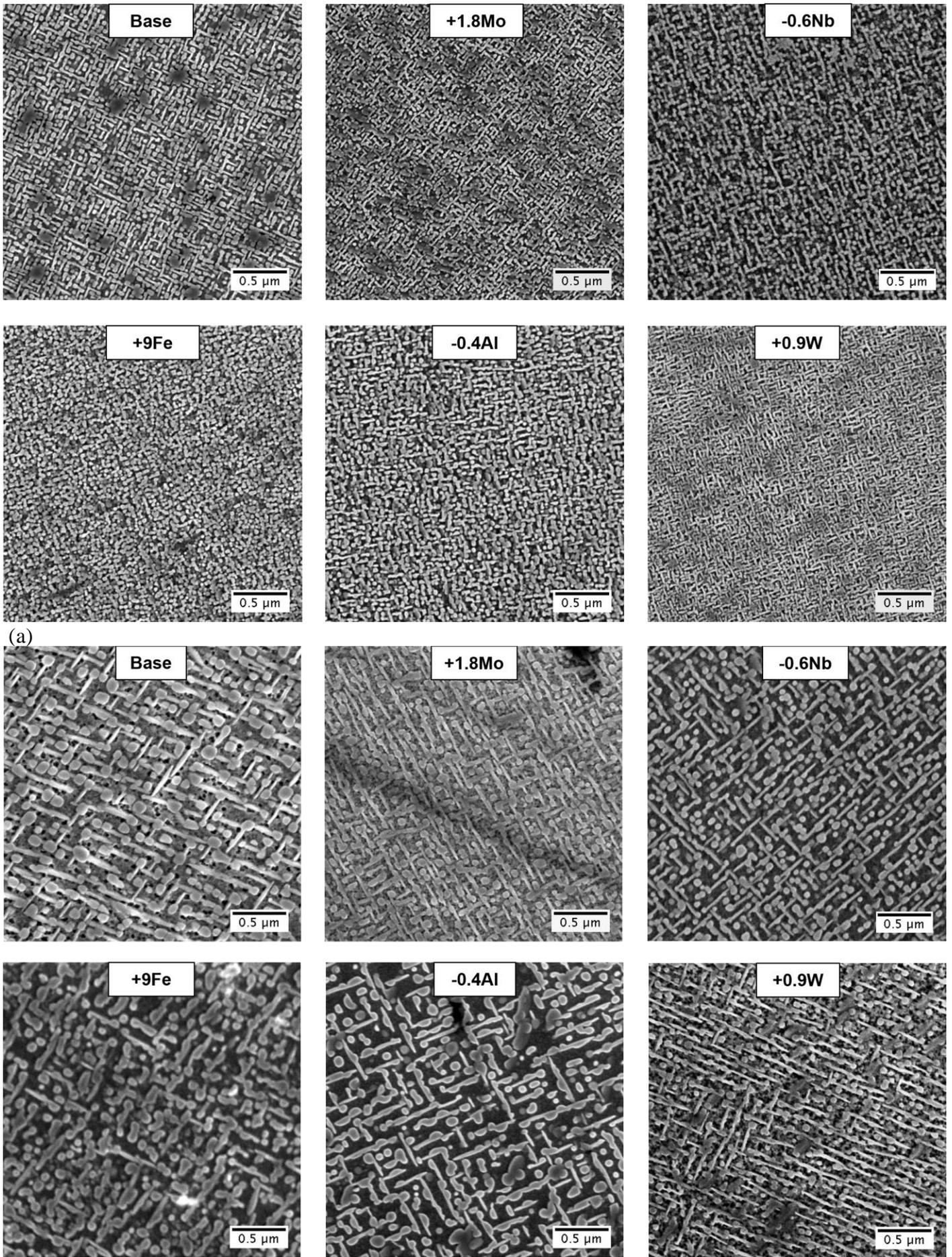

*Figure 5: In-lens SEM images of the microstructure of each alloy after (a) 10-hour heat treatment and (b) 100 hour heat treatment at 800 °C.*

Atom probe tomography was utilized to measure accurate chemical compositions and assess the effect of individual alloying additions on partitioning behavior in the γ′ and γ″ phases. The samples heat treated for 10 hours were chosen for this study because the finer microstructure would allow detection of multiple phases within each atom probe needle. In comparison to the SEM analysis, APT allows the unambiguous identification of the precipitate phases by their compositional differences.

An atomistic scale reconstruction of the APT sample of the base alloy is shown in Figure 6(a) in which the γ, γ′ and γ″ phases may be separately distinguished. Due to the complex shapes and number of phases present, proximity histograms could not be utilized to analyze their interfaces. Instead, regions of interest (50x20x20 nm) were selected and located perpendicular to the interface to be analyzed. This method, which provides 1-D composition profiles, is very effective for measuring chemical compositions at the interfaces, although if the selected area is not perfectly parallel to the interface, its width can be artificially inflated. Figure 6(b) shows a 1-D composition profile going through all the phases present. The solid solubility of Nb in the γ′ phase is substantially higher than that of Al in γ″, which is consistent with previous semi-quantitative EDX observations [13].

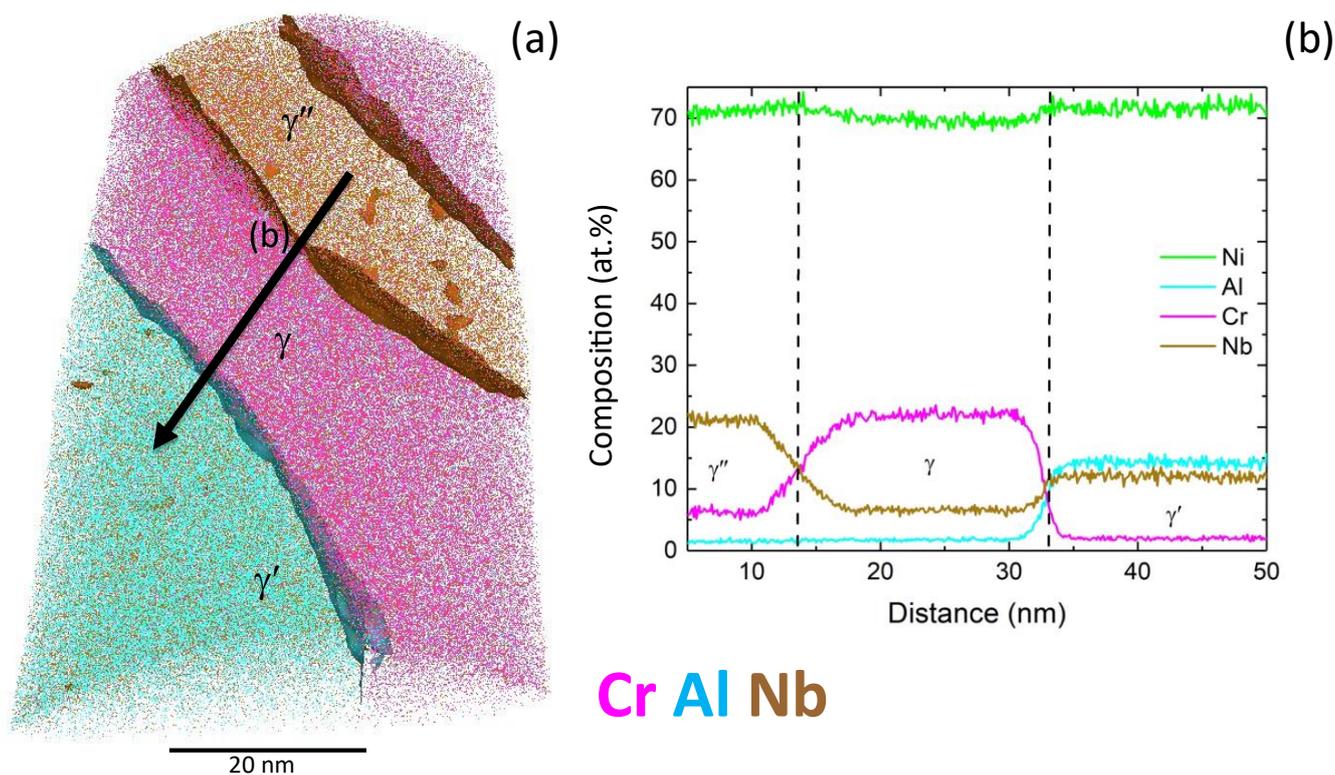

*Figure 6: (a) Atom probe tomography reconstruction of the base alloy, showing Cr, Al, and Nb atom distributions in the three phases γ, γ′, and γ″, separated by iso-concentration surfaces of 7 at.% Al and 15 at.% Nb; (b) 1-D composition profile going across all the phases present, the location the composition profile was taken from is shown in Figure 6(a) as a white arrow.*

The alloy with 1.8 at% Mo was also analyzed by atom probe tomography and a reconstruction is shown in Figure 7(a), where the cuboidal precipitate morphology of the γ′ precipitates can be observed.

One-dimensional composition profiles across the γ-γ′-γ″ phase interfaces are shown in Figure 7(b) to (d). – The added Mo is predominantly accommodated within the γ phase as well as in the γ″ phase but is absent in the γ′ phase. Figure 7(d) reveals a depletion in Ni at the γ′-γ″ interface. While the interface between the γ and the γ′ phase is clear, the interface between the γ and γ″ phases as well as between the γ′ and γ″ phases (Figure 7(c) and (d)) are diffuse, which can be observed through the "serrations" in the iso-concentration surfaces shown in Figure 7(a). The serrations are likely to be an artefact of the data binning algorithm.

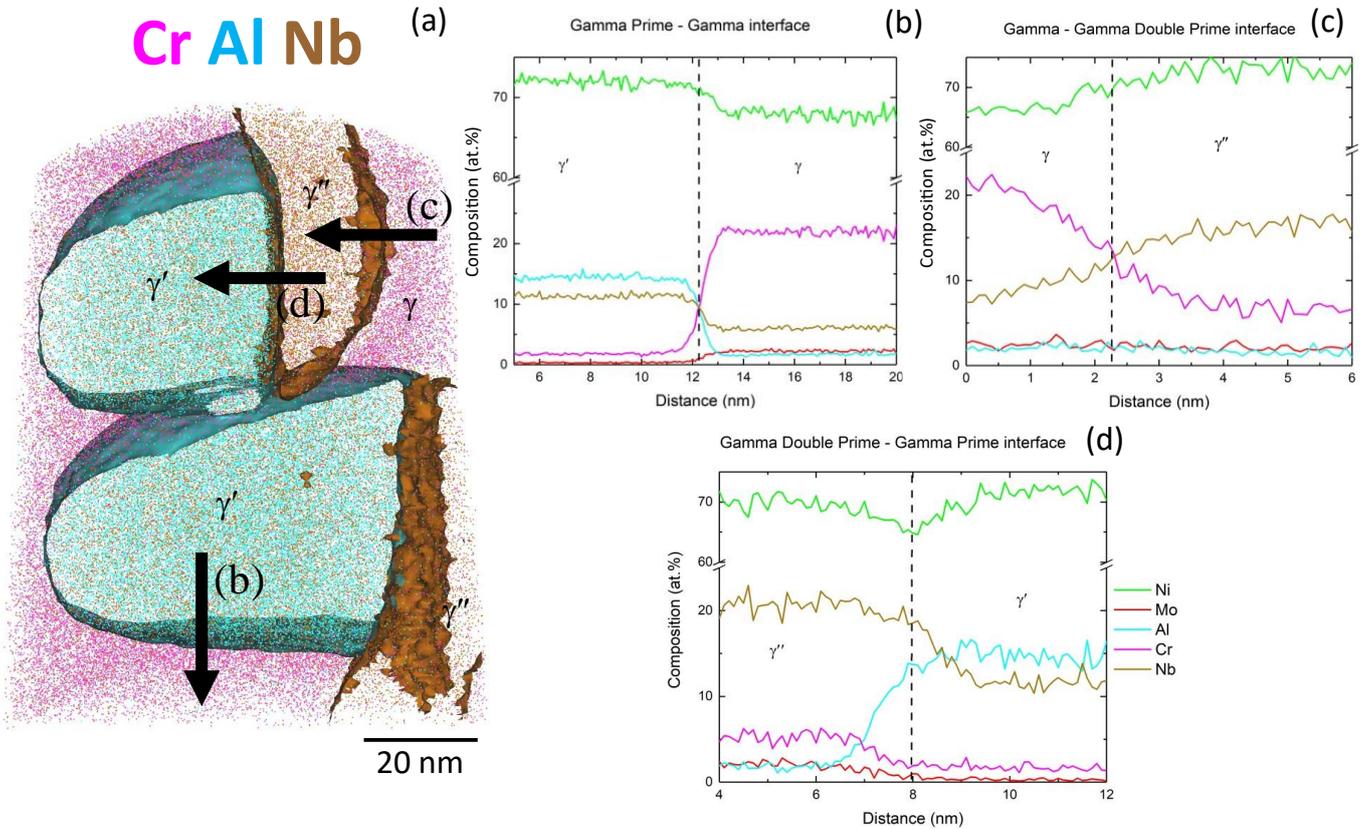

*Figure 7: (a) APT reconstruction of a specimen of the alloy with 1.8 at.% Mo, iso-concentration surfaces of 7 at.% Al and 15 at.% Nb envelope the precipitate phases γ′ and γ″; (b) detailed view of crystallographic zone axis from the reconstruction shown in (a); (c), (d), and (e)) 1-dimensional composition profiles across all phase interfaces from the reconstruction shown by arrows in (a).*

The 3D reconstruction of the -0.6Nb APT specimen is shown in Figure 8 (a). The lenticular γ″ shape is clearly seen in this specimen. It is also noteworthy that the γ″ precipitate has grown around a γ′ precipitate, indicating that the γ′ precipitates formed first. 1D composition profiles through all three phases are presented in Figure 8 (b-d). No apparent differences to the previous alloy are observed from the reduction of Nb, with the only exception that the γ′-γ″ interface is slightly enriched in Ni, whereas the same interface was depleted in Ni for the previous alloy. The alloying alterations may have caused a change in interfacial energy of the precipitate phase interfaces, leading to decoration or expulsion of Ni at the interface.

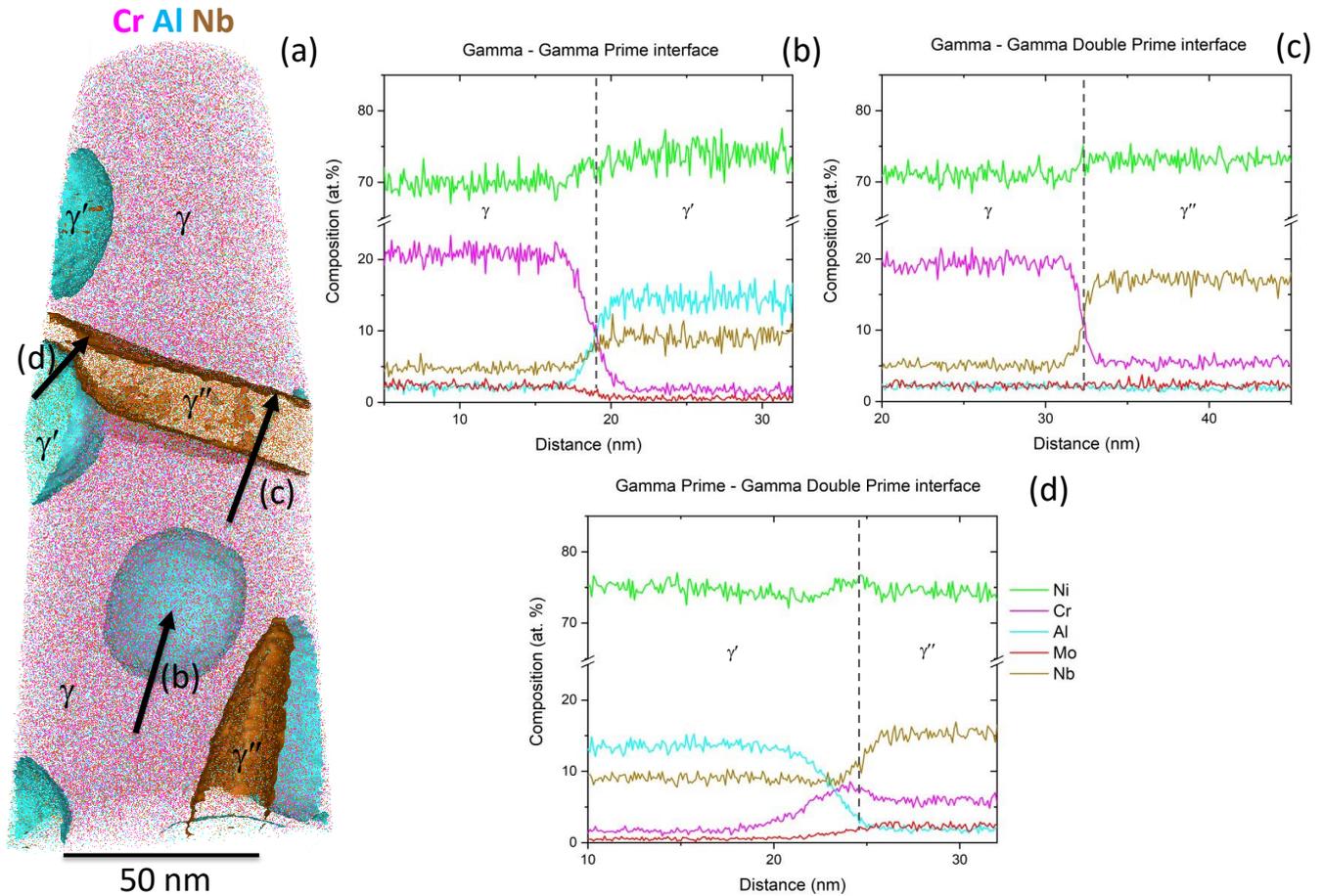

*Figure 8: (a) APT reconstruction of a specimen of the -0.6Nb alloy with iso-concentration surfaces of 7 at.% Al and 13 at.% Nb, showing that the Nb-rich γ″ phase has grown around an Al-rich γ′ precipitate, (b), (c), and (d) 1-dimensional composition profiles across all phase interfaces from the reconstruction shown in (a).*

The APT reconstruction of the +9Fe alloy is shown in Figure 9(a). Based on the SEM analysis shown in Figure 5, the presence of Fe coarsened the microstructure and led to more spherical than cuboidal γ′ phase precipitates which have grown into one another as well as into the γ″ phase. The spherical shape of the γ′ precipitates and their merging is also observable in the APT reconstruction, as can be seen for the irregular precipitate in the back of the reconstruction, likely having merged from two or three γ′ phase precipitates.

Inspection of the interfacial composition profiles, shown in Figure 9(b) to (d), indicate that Fe segregated preferentially to the γ phase, although low levels were found in solid solution both in the γ′ and in the γ″ phases. In contrast to the +1.8Mo alloy but similarly to the -0.6Nb alloy, there is an excess nickel segregation at the γ′-γ″ interface.

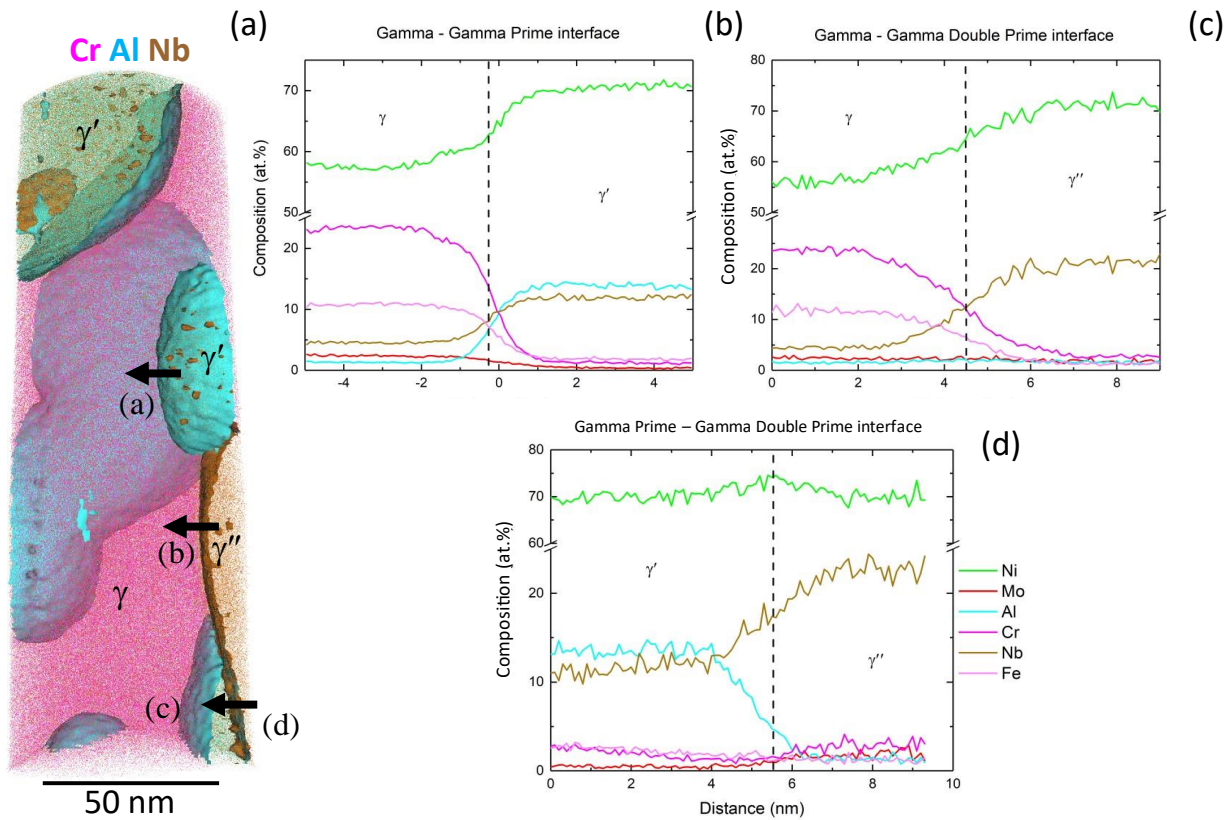

*Figure 9: (a) APT reconstruction of the +9Fe alloy containing 9 at.% Fe with iso-concentration surfaces of 7 at.% Al and 15 at.% Nb, (b) proximity histogram showing the elemental composition profile across the γ-γ′ interface, (c) 1-D elemental composition profile across the γ–γ″ interface and (d) 1-D elemental composition profile across the γ′-γ″ interface. The locations the figures (b-d) were taken from is highlighted by arrows in Figure (a).*

Figure 10 (a) shows an APT reconstruction of the -0.4Al alloy. The lenticular nature of the γ″ phase is, again, clearly seen and the γ′ phase is found to sit on the side of the γ″ phase in accordance with the observations made from the SEM micrographs. One γ′ precipitate is found sitting partially within the γ″ phase. This is an observation that the SEM had not been able to provide due to the lack of contrast between the precipitate phases.

Figure 10(b), (c) and (d) show 1-dimensional composition profiles across phase interfaces. There are no apparent differences in the phase compositions and interface segregation of this alloy compared to the +9Fe alloy.

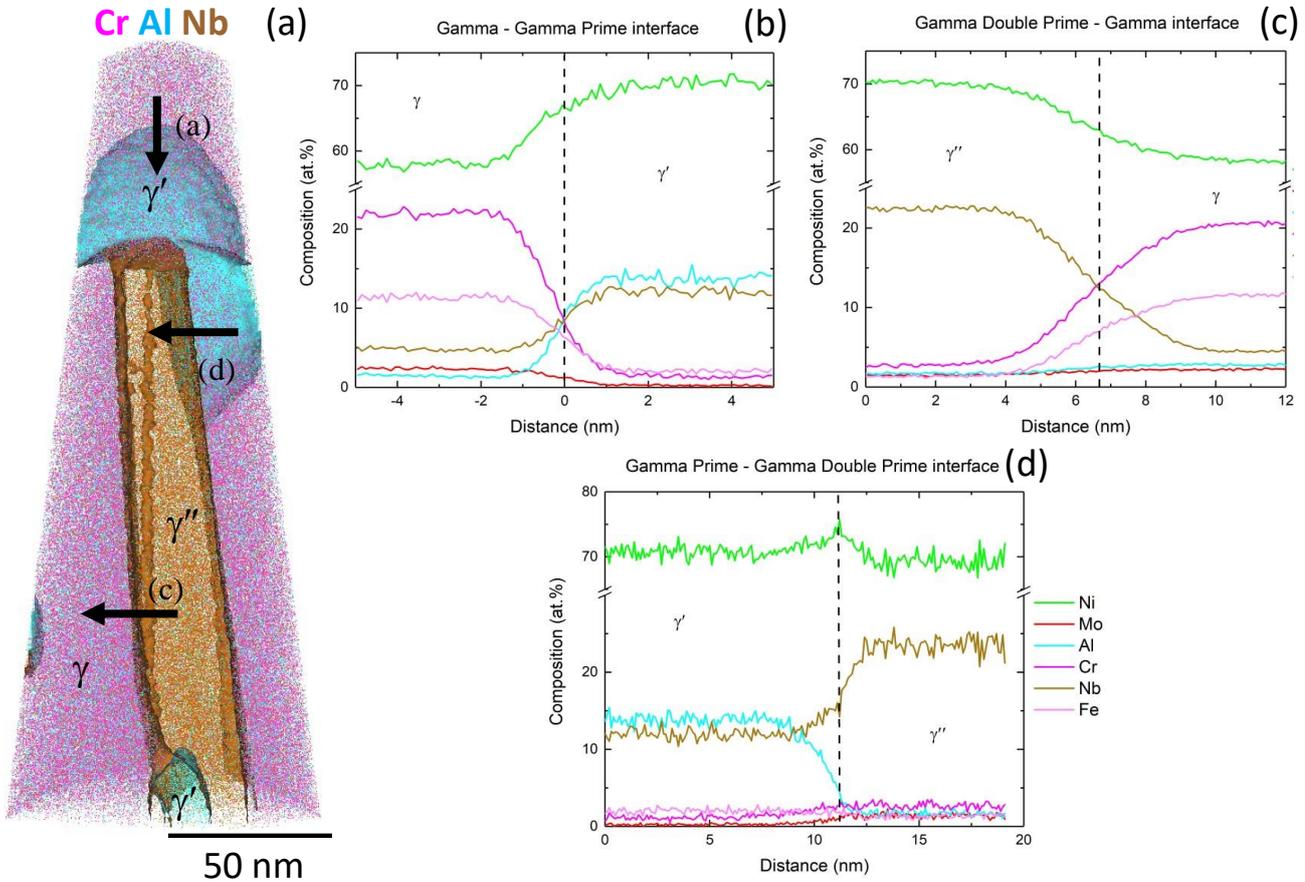

*Figure 10: (a) APT reconstruction of the -0.4Al alloy with iso-concentration surfaces of 7 at.% Al and 15 at.% Nb, (b) proximity histogram showing the elemental composition profile across the γ-γ′ interface, (c) 1-D elemental composition profile across the γ″-γ interface and (d) 1-D elemental composition profile across the γ′-γ″ interface. The locations of the 1-D composition profiles are highlighted by arrows in (a).*

Figure 11(a) shows an APT reconstruction of the alloy with 0.9 at% W. The +0.9W alloy does not contain Fe and exhibited phase compositions comparable to the -0.6Nb alloy apart from low levels of W in each phase. W partitioned preferentially to the γ″ phase (Figure 11(d)), although low-levels (<0.5 at.%) were detected within the γ and γ′ phases as well (Figure 11(b) and (c)). In contrast to the previous alloys, the γ′-γ″ interface exhibits neither a depletion nor an enrichment of Ni.

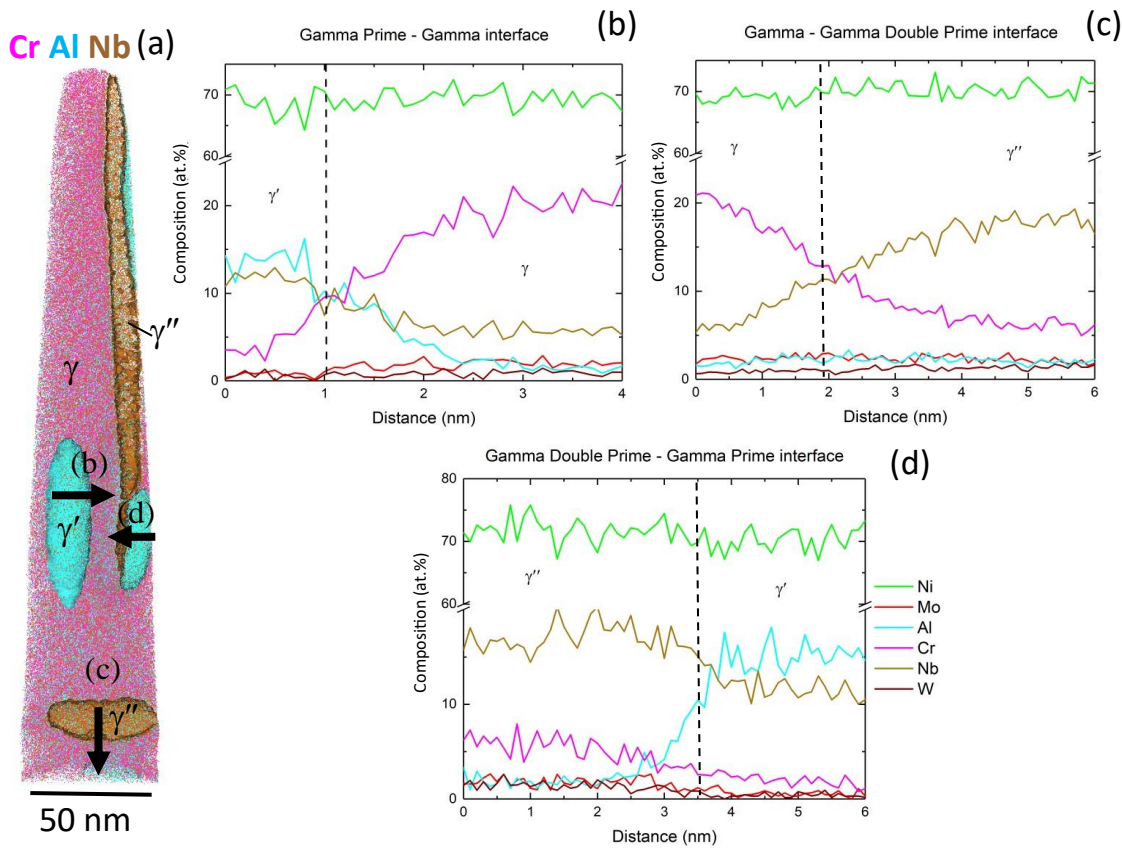

*Figure 11: (a) atomic scale reconstruction of the alloy containing 0.9 at.% W and no Fe, with iso-concentration surfaces of 7 at.% Al and 15 at.% Nb, (b) (c) and (d) 1-D elemental composition profiles across interfaces within the reconstruction. The location of the 1-D concentration profiles is shown in figure (a) in the form of arrows.*

For accurate elemental compositions of the γ, γ′ and γ″ phases, peak decomposition analysis was performed on the atom probe tomography data. This analysis yields more precise phase compositions than the 1-D elemental composition profiles, which are used to analyze the interfaces. The decomposition analysis was performed on spectra from isolated single-phase volumes, which were delineated with the iso-concentration surfaces show in in Figure 11, with the interfacial region excluded due to its data not being representative. This was done by exporting the individual phases as a separate file, then selecting a central region of interest (ROI) within them. The peak decomposed APT compositions are shown in Figure 12(a), (b) and (c), respectively (values also given in appendix Table AIII to Table AV) From this, conclusions can be drawn about the effects of the alloying additions on the micro-segregation of the elements. The alloying additions changed the composition as follows:

**+1.8Mo:** With addition of 1.8 at.% Mo, some compositional changes can be observed compared to the base alloy. In the matrix γ phase of the +1.8Mo alloy, Mo took up 2.5 at.%, which is consistent with a reduction of Ni by 2.5 at.% from the value in the base alloy. This suggests that in the matrix γ phase Mo merely substituted for Ni. In the γ′ phase, the presence of Mo had no apparent effect on the composition, although 0.3 at.% Mo was found in solid solution. Mo, however, triggered dramatic changes to the γ″ phase composition by decreasing the Al and increasing the Cr solid solubility. In the base alloy, γ″contained 6.3 at.% Al and 1.5 at.% Cr, whereas the +1.8Mo alloy contained 5.9 at.% Cr and 1.8 at.% Al. Also, the Nb content was reduced by 2 at.% and Mo was found at 2.2 at.% in solid solution.

**-0.6Nb:** As the Nb content was reduced from 6 to 5.4 at.%, the γ phase exhibited a slight increase in Ni (+ 1.7 at.%) and a slight decrease in Cr (- 1.3 at.%). The Nb reduction does not seem to have affected the elemental segregation in the precipitate phases γ′ and γ″.

**+9Fe:** The highest content of Fe was found in the γ phase, where it mainly substituted for Ni (Ni -12.8 at.%, Fe +12.4 at.%) and also resulted in a slight increase of Cr by 1.5 at.% along with a reduction of Nb by 0.9 at.%. Only 2 at.% Fe was found in the the γ′ phase and the Al content was reduced by 1 at.%. More dramatic effects were observed for the γ″ phase where the solubilities of Nb and Cr were clearly affected. Here, the Nb content increased by 4 at.% and Cr decreased by 3.3 at.%. Fe was present in a low amount of 1.5 at.%.

**-0.4Al:** The 10% reduction of Al did not greatly affect the phase compositions. Only in the γ phase, there were slight variations in the segregation behavior of Ni and Cr: the Ni content increased by 1.4 at.% while Cr decreased by 1.5 at.%. In the precipitate phases, there were no visible compositional differences to the +9Fe alloy.

**+0.9W:** The +0.9W alloy, which does not contain Fe, did not exhibit any strong compositional changes. W was found in all phases with a preferential segregation to the γ″ phase (γ: 0.8 at.%, γ′: 0.4 at.%, γ″: 1.4 at.%). This alloy had the highest γ′ phase Al content (15.1 at.%) out of all the investigated alloys and the lowest γ″ phase Nb content (15.3 at.%).

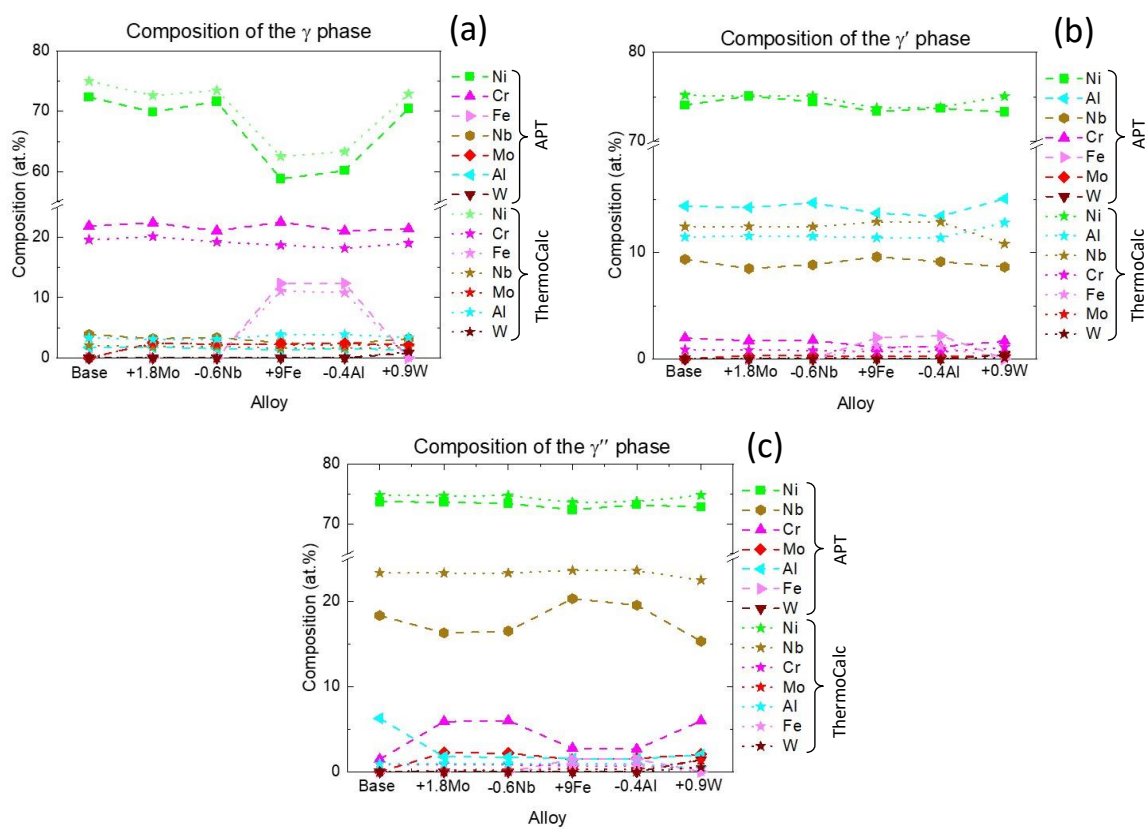

*Figure 12: Elemental composition of the (a) γ, (b) γ′ and (c) γ″ phase for all investigated alloys as measured by APT using peak decomposition versus their respective Thermo-Calc predictions. APT data is represented by the dashed lines and the symbols according to the legend); Thermo-Calc data is represented by the respectively colored dotted lines with star symbols.*

Figure 12 also shows the phase compositions as predicted by Thermo-Calc at 800˚C. For the γ matrix, the predicted data matched the experimental data well with only slight deviations, such as the predicted Ni content being slightly above the measured data but following the trends and the Cr prediction being slightly below the measured content. While the prediction of Ni content in the γ′ phase matches the experimental data very well, larger deviations are found for the two precipitate forming elements Al and Nb. It worth noting, that this is likely related to the observed temperature offset in the Thermo-Calc predictions, as has been observed in other studies [33 - 35] , The comparison of the γ′ solvus temperature as measured by DSC and as predicted by Thermo-Calc (Figure 2) revealed an offset of approximately -70 °C. Thermo-Calc predictions of the phase compositions calculated at an isothermal temperature of 700 °C, i.e. at a temperature which compensates somewhat for the observed offset, are in very good agreement to the measured compositions for the γ and γ′ phase, see appendix Figure A1. For both the compositions calculated at 800 °C shown in Figure 12 and those calculated at 700 °C, the Thermo-Calc predictions for the γ″ phase, showed large discrepancies from the experimentally obtained data. The greatest deviation was found for Nb, followed by Cr, which is likely due to difficulties in performing calculations involving metastable phases. The deviations may also be a reflection of the fidelity of the thermodynamic databases across this particular region of composition space.

## 4. Discussion

Compositional variations to a promising polycrystalline Ni-based superalloy have been investigated to get insights into the role of alloying elements and to further promote the development of new and improved superalloys for turbine disc applications. One of the questions addressed in this work is whether the beneficial properties of the base alloy, i.e. a high temperature stability and yield strength, can be retained in the compositionally modified alloys. These and other properties are discussed in the following.

### 4.1 Alloying effects on temperature stability and hardness

The γ′ solvus temperature was determined for all alloys. Both precipitate forming elements, i.e. Al and Nb, were found to have a considerable positive effect on the solvus temperature. It can hence be deducted that these elements not only lead to the formation of γ′ but also stabilize this phase at elevated temperatures. The other alloying elements (Mo, W, and Fe) were found to decrease the solvus temperature with a lower absolute potency, Mo having a negligible effect.

Effects of alloying additions on the γ′ solvus temperature have been reported in literature before. Mo, for example has been shown to increase γ' solvus by 2.5 °C/at.%, despite partitioning to the γ phase primarily [19], whereas W has been shown to increase γ' solvus ~ 4 °C/at.% [30], contrary to our findings. Both studies analyzed SX alloys without Nb, but with B, C, Hf, Y/Zr, additionally Co, Re/Ti. It is possible that one of the other differences in composition is responsible for this discrepancy. These opposing results show the complexity of the interplay of alloying additions and preparation methods. An alloying element may have opposing effects depending not only on the content of this element but also on the total composition as the various elements may all affect each other. Additionally, the temperature of the applied heat treatments can have a significant impact on the overall material characteristics and properties. To address this, thermodynamic modelling using Thermo-Calc is useful for exploring the trends in γ′ solvus temperatures, but the results obtained suggest that care must be taken when using these models for alloy design due to some of the discrepancies in absolute values.

XRD results from the samples heat treated at 800°C revealed the presence of the γ′, γ″ and δ phases in all alloys. To give precise information about the thermal stability of this set of alloys with regards to the γ″ to δ phase transformation, further studies are needed at lower temperatures such as 700 and 750 °C. However, this is beyond the scope of the current work.

The yield strength of the alloys was not measured directly but instead hardness data were used as a proxy, allowing alloying effects to be investigated. One key question this work wanted to answer was whether the yield strength could be retained when reducing the Nb and Al content to achieve an improved processability. Such a reduction would reduce the precipitate phase fraction, which may lead to a reduced hardness. Our data shows that the Al reduction does not notably affect the alloy hardness, whilst the Nb reduction significantly reduces the alloy hardness. This effect, which may be putatively attributed to a decreased precipitate fraction, but also due to decreased solid solution strengthening, should be considered for further alloy development.

Mo and W, which are both added for solid solution strengthening, have a positive effect on the hardness, which confirms their role as solid solution strengtheners. This effect of Mo and W has also previously been reported on individual rather than successive alloying additions to the Ni-15Cr-4Al-6Nb (at.%) system [13].

## 4.2 Alloying effects on microstructure and partitioning

SEM micrographs show that Mo has a refining effect on the microstructure, and Fe additions result in precipitate coarsening and clustering, such that γ′ and γ″ precipitates can no longer be clearly distinguished. While Nb was reported to reduce precipitate coarsening [24], the slight reduction of 0.6 at.% does not seem to have a strong effect on the microstructure besides reducing the precipitate volume fraction, similar to the reduction of Al. Comparing this with previous work by Mignanelli *et al.*[14], they showed that Mo and Fe had no visible effect on the microstructure when observed by SEM, whereas W lead to a "loss of morphological distinction". However, in their study alloying elements were added individually (to different alloy samples), whereas in our study, these elements were added successively, which may account for the discrepancies in observations due to the interplay between elements.

Due to insufficient contrast between the precipitate phases, phase volume fractions could not be retrieved from the SEM micrographs. However, it is possible to derive the phase fractions from the APT data, from which compositions for all three principal phases were obtained. As these compositions need to add up to the overall sample composition, a linear equation system can be set up. Due to uncertainties in the compositional values, a solution for the phase fraction can be found by solving a constrained least square problem. Two constraints were applied: 1. all phase fractions need to be between zero and one and 2. all phase fractions must add up to one. The so derived atomic phase fractions can be converted to volume phase fractions with the knowledge of the volume per unit cell, taken from the XRD measurements, and the number of atoms per unit cell as given by the well-known crystal structures of the phases.

The changes in composition near interfaces are not considered in this calculation and can hence lead to some deviations from the actual volume fractions. However, the so derived data, presented in Figure 13, can be used for relative comparisons between the different alloys and hence can be related to alloying effects of the individual alloying elements. The presence of the δ phase in all alloys is not considered to have a noticeable influence on this analysis, as the δ phase was only found in small quantities at the grain boundaries, away from the lift-out regions for the atom probe needles.

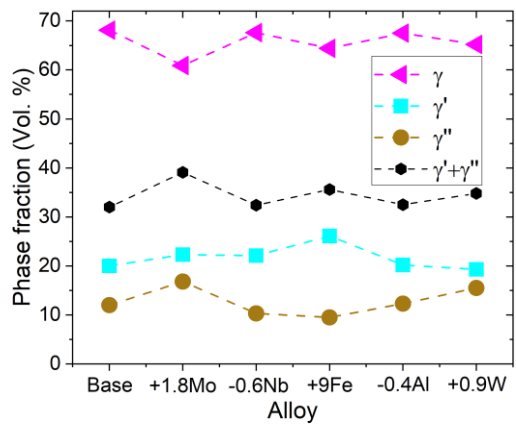

*Figure 13: Phase volume fractions of all alloys as derived from APT phase compositions.*

It has previously been mentioned that the Al/Nb ratio is indicative of the type of phase precipitation in a Ni-based superalloy, with ratios larger than 1 leading to two-phase systems of γ and γ′ only [36] and ratios below 0.3 leading to γ and γ″. Therefore, the base alloy was developed with an intermediate ratio in order to achieve the coexistence of all three phases [13]. For all alloys in this work, the Al/Nb ratio is between 0.67 and 0.74, i.e. intermediate, and all contain the three phases γ, γ′ and γ″. The alloys with a larger ratio show also a reduced γ″ fraction which agrees with the above statement, that larger Al/Nb ratios lead to the disappearance of γ″.

To assess the effect of individual alloying additions and alterations, it is helpful to consider the evolution of the phase volume fractions in conjunction with elemental partitioning and the other results presented.

### 4.3 Effect of Mo

Upon addition of Mo to the base alloy, both precipitate volume fractions increase, though γ″ does so more strongly than γ′. The matrix phase volume fraction is reduced accordingly. Looking at the elemental partitioning, Mo is found to partition preferentially to γ and γ″ (>2 at.% vs <0.4 in γ′). It also leads to a change in Al partitioning in the γ″ phase. Al is strongly expelled from the γ″ phase, and also the Nb content is reduced by Mo while Cr is enriched. The reduction of the Nb composition needed for γ″ formation may explain, in part, the increase in γ″ volume fraction. The expelled Al can instead go to γ′ which explains the increase in γ′ phase fraction.

The +1.8Mo alloy is found to have the highest total precipitate fraction (γ′ + γ″) of all alloys. As the SEM micrograph shows a refined microstructure, it can be concluded that the addition of Mo leads to the precipitation of a finer and denser dual superlattice microstructure.
This change in microstructure and the increase in total precipitate fraction may contribute to the slight increase in hardness observed, additionally to a solid solution strengthening mechanism.

### 4.4 Effect of Nb

Nb partitions to both precipitate phases, but more strongly to the γ″ phase (around 15-20 at. % vs. 8-10 at. % in γ′). Upon reduction of Nb, the γ″ volume fraction is strongly reduced in favor of the matrix phase γ, while the γ′ volume fraction stays the same. This is consistent with Nb being a strong γ″ former. It is, however, also one of the major constituent elements of γ′ but clearly does not determine the γ′ volume fraction. The fact that the γ′ volume fraction is not affected by the reduction in Nb leads to the conclusion that Nb first occupies the γ′ precipitates and only when γ′ is saturated in Nb, the remaining atoms form γ″. This is supported by the fact that γ′ forms prior to γ″ as has previously been shown through TTT diagrams [REF], and as is suggested by the present atom probe reconstructions in Figures 8, 10, 11. It is also supported by the previously mentioned relationship between Al/Nb ratio and prevalent precipitate phases [13] (see also section 4.2),

When reducing the Nb composition, due to the reduced γ″ volume fraction, the total precipitate fraction is again at a similar level to the base alloy. The hardness, however, is strongly reduced compared to both the Base and the +1.8Mo alloy. This could be due to reduced solid solution strengthening with reduced Nb content. However, the observation could also suggest that that γ″ may be a more effective precipitate strengthener than γ′.

### 4.5 Effect of Fe

As Fe is added for improved processability and reduced material cost, it is to be assessed whether the additions lead to any deteriorations in superalloy properties. Fe can trigger precipitation of detrimental Laves phases. It is a promising result that these phases were not found in the present alloys.

Regarding elemental partitioning to the γ-γ′-γ″ phases, Fe clearly partitions to the matrix γ-phase, albeit with low levels found in both the γ′ and γ″ precipitate phases. While Fe was reported to aid in γ″ formation [25], in our three-phase system it instead leads to an increase in the γ′ volume fraction and a reduction in the matrix volume fraction (and slight reduction in γ″). In the case of γ′-γ″ co-existence, Fe additions may aid in the formation of γ′, even though the γ′ solvus temperature is decreased. Further studies are needed to investigate underlying mechanisms. In the γ″ phase, the addition of Fe increases the solubility of Nb and decreases the Cr solubility. Excess Ni segregation is observed at the γ′- γ″ interface for both alloys containing Fe. A coarsening of the microstructure and precipitate clustering is observed.

## 4.6 Effect of Al

Al partitions strongly to the γ′ phase and a reduction in Al leads to a substantial reduction in the γ′ volume fraction and concomitant increases in both γ and γ″ fractions. Clearly, Al is a strong γ′ former with no apparent additional effects on partitioning or phase precipitation.

## 4.7 Effect of W

W is found to partition preferentially to the γ″ phase and is least available in γ′, which is a differing observation to that of Amouyal et al. [18]. As previously mentioned, they had shown that in γ-γ′ containing Ni-superalloys, W preferentially occupies Al sites in the γ′ phase unless driven out into the matrix by Ta additions. The alloys in this work do not contain Ta, although they do contain Nb and γ″ as a second precipitate phase. The radius of W is close to that of Nb, both of which are larger than the radius of Al. It is hence not surprising that W is found preferentially in the γ″ phase, likely substituting for Nb. Taking into account the effect of Fe removal on phase fractions, W leads to the increase of both precipitate volume fractions and hence aids in the formation of precipitates. Critically, neither the addition of W nor the addition of Mo at the concentrations considered have led to precipitation of detrimental TCP phases during the thermal exposures performed in this work.

## 5. Conclusions

In this study, we have assessed the effect of individual alloying additions Mo, W, Fe and variations of Nb, and Al on the properties of a new dual-superlattice reinforced nickel-based superalloy. Thermal stability, hardness, elemental partitioning, microstructure and phase fractions were investigated. The results presented lead to the following conclusions:

1. The addition of 1.8 at.% Mo does not significantly decrease the γ′ solvus temperature and prevents microstructural coarsening during heat treatments. A slight increase in hardness is observed, which can be attributed to solid solution strengthening and an increased precipitate volume fraction, i.e. increased precipitate strengthening. The alloying goal (increased strength) has hence been reached without having any apparent negative effect on the alloy.
2. A reduction of the overall Nb content by 10 %, i.e. 0.6 at.% reduction, strongly reduces the γ″ volume fraction, without affecting the γ′ volume fraction. The reduced precipitate fraction leads to a significant reduction in alloy hardness. Reductions of Nb for improved processability are hence to be handled cautiously, given its effects on superalloy strength.
3. A reduction of the overall Al content by 10 %, i.e. 0.4 at.% reduction, reduces the γ′ volume fraction and effectively reduces the γ′ solvus temperature, but does not have an apparent effect on the alloy hardness. Both Al and Nb are confirmed as γ′ and γ″ forming elements, respectively, and found to stabilize the γ′ phase to increasing temperatures. Reductions of Al to improve alloy processability may hence be preferrable to reductions of Nb, although the potency of Al to reduce the γ′ solvus temperature needs to be kept in mind.
4. Fe, substituted with Ni for better processability and reduced material cost, decreases the γ′ solvus temperature and coarsens the microstructure during heat treatment but does not have an apparent effect on the alloy hardness. It is found mostly in solid solution in the γ phase.
5. The addition of 0.9 at.% W decreases the γ′ solvus temperature but increases the precipitate volume fractions, which in addition to solid solution strengthening causes an increase in alloy hardness, which was the aim of this alloying addition.
6. Thermo-Calc predictions of γ′ solvus temperatures capture some of the effects of compositional variation, but with an offset of approximately 70 °C. Thermo-Calc predictions on elemental partitioning can also capture effects of compositional variation, albeit with varying precision depending on alloying element and phase. Thermo-Calc predictions using an isothermal temperature of 700 °C, that account somewhat for the observed temperature offset, have provided improved agreement of predicted and experimentally obtained data.

Further to this study, it will be useful to investigate the role of alloying on the formation of the δ phase. This is part of ongoing work.

## Acknowledgements

Funding for this project was provided by the Rolls-Royce/EPSRC strategic partnership grant, EP/M005607/1. S. Pedrazzini acknowledges EPSRC EP/S013881/1 and the RAEng's support by means of an associate research fellowship. The LEAP 5000XR used in this study was funded by EPSRC Grant (No. EP/M022803/1).

*Conflict of interest:* The authors declare that they have no conflict of interest.

# Appendix

The compositions of all alloys were measured by doing five large area EDX maps on the alloys, the averaged data is presented in Table AI.

*Table AI: Alloy compositions in at. % as averaged over five large EDX maps.*

| Composition (at. %) | Ni | Cr | Nb | Al | Mo | Fe | W |
|---|---|---|---|---|---|---|---|
| Base | 73,0 | 16,7 | 6,3 | 4,0 | 0,0 | 0,0 | 0,0 |
| +1.8Mo | 72,3 | 16,2 | 5,8 | 3,9 | 1,8 | 0,0 | 0,0 |
| -0.6Nb | 73,8 | 15,5 | 5,2 | 3,9 | 1,6 | 0,0 | 0,0 |
| +9Fe | 64,6 | 15,6 | 5,1 | 3,8 | 1,6 | 9,3 | 0,0 |
| -0.4Al | 62,9 | 16,8 | 5,2 | 3,6 | 1,9 | 9,6 | 0,0 |
| +0.9W | 72,8 | 16,4 | 5,2 | 3,4 | 1,8 | 0,0 | 0,4 |

The γ' solvus temperature was retrieved from the DSC thermograms at the inflection point of the S-shaped curve at around 1000 °C by curve differentiation and peak identification. The so derived γ' solvus temperatures are listed in Table AII.

*Table AII: γ' solvus temperature as obtained from DSC data.*

|  | $T_{solvus}$ (°C) |
|---|---|
| Base | 1010 |
| +1.8Mo | 1009 |
| -0.6Nb | 1005 |
| +9Fe | 988 |
| -0.4Al | 970 |
| 0.9W | 983 |

Compositions of the γ, γ', and γ'' phases were taken from APT data by peak decomposition of spectra taken from isolated single-phase volumes and are presented in Table AIII to Table AV.

*Table AIII: Compositions of the γ phase in at. % as obtained from APT decomposition data.*

| Composition (at. %) | Ni | Cr | Nb | Al | Mo | Fe | W |
|---|---|---|---|---|---|---|---|
| Base | 72,37 | 21,86 | 3,88 | 1,73 | 0 | 0 | 0 |
| +1.8Mo | 69,88 | 22,39 | 3,22 | 1,76 | 2,51 | 0 | 0 |
| -0.6Nb | 71,62 | 21,06 | 3,37 | 1,57 | 2,2 | 0 | 0 |
| +9Fe | 58,81 | 22,51 | 2,5 | 1,35 | 2,32 | 12,36 | 0 |
| -0.4Al | 60,16 | 21,02 | 2,35 | 1,54 | 2,44 | 12,39 | 0 |
| 0.9W | 70,42 | 21,39 | 3,19 | 0 | 2,14 | 1,32 | 0,8 |

*Table AIV: Compositions of the γ' phase in at. % as obtained from APT decomposition data.*

| Composition (at. %) | Ni | Cr | Nb | Al | Mo | Fe | W |
|---|---|---|---|---|---|---|---|
| Base | 74,06 | 1,97 | 9,39 | 14,41 | 0 | 0 | 0 |
| +1.8Mo | 75,06 | 1,72 | 8,52 | 14,28 | 0,3 | 0 | 0 |
| -0.6Nb | 74,43 | 1,75 | 8,87 | 14,71 | 0,32 | 0 | 0 |
| +9Fe | 73,32 | 1,09 | 9,63 | 13,74 | 0,21 | 2 | 0 |
| -0.4Al | 73,7 | 1,14 | 9,17 | 13,45 | 0,24 | 2,2 | 0 |
| 0.9W | 73,3 | 1,68 | 8,66 | 15,1 | 0,34 | 0 | 0,4 |

*Table AV: Compositions of the γ'' phase in at. % as obtained from APT decomposition data.*

| Composition (at. %) | Ni | Nb | Cr | Al | Mo | Fe | W |
|---|---|---|---|---|---|---|---|
| Base | 73,66 | 18,36 | 1,46 | 6,28 | 0 | 0 | 0 |
| +1.8Mo | 73,64 | 16,32 | 5,89 | 1,76 | 2,24 | 0 | 0 |
| -0.6Nb | 73,4 | 16,54 | 6,01 | 1,72 | 2,18 | 0 | 0 |
| +9Fe | 72,36 | 20,34 | 2,76 | 1,55 | 1,45 | 1,46 | 0 |
| -0.4Al | 73,18 | 19,58 | 2,65 | 1,49 | 1,56 | 1,47 | 0 |
| +0.9W | 72,77 | 15,34 | 5,98 | 2 | 2,07 | 0 | 1,4 |

Figure A1 shows the composition of the three phases γ, γ', and γ'' as measured by APT on samples heat treated at 800 °C and as predicted by Thermo-Calc assuming an isothermal temperature of 700 °C. This difference in temperature somewhat compensates for a temperature offset observed for Thermo-Calc predictions.

For the γ matrix and the γ′ precipitate phase, the predicted data matched the experimental data well with only slight deviations. For the γ phase, the predicted Ni content was slightly above the measured data but follows the trends and the Cr prediction was slightly below the measured content of the Fe-free alloys. For the γ′ phase, the predicted Al content was slightly below the measured content of the Fe-free alloy. The Ni predictions showed some excess in the base alloy and the +0.9W alloy but otherwise coincided very well with the experimental data. For the γ″ phase, however, the Thermo-Calc predictions showed large discrepancies from the experimentally obtained data. This is likely due to difficulties in performing calculations with metastable phases, but also may have occured due to the fidelity of the thermodynamic databases across this particular composition space. The greatest deviation was found for Nb, followed by Cr. Comparing the different alloys, the Thermo-Calc predictions deviated most from the experimental data for the alloy containing W. This was especially visible for Ni and Al in γ′ and for Nb and W in γ″. The best agreement between the predicted and measured compositions was achieved with the Fe-containing alloys.

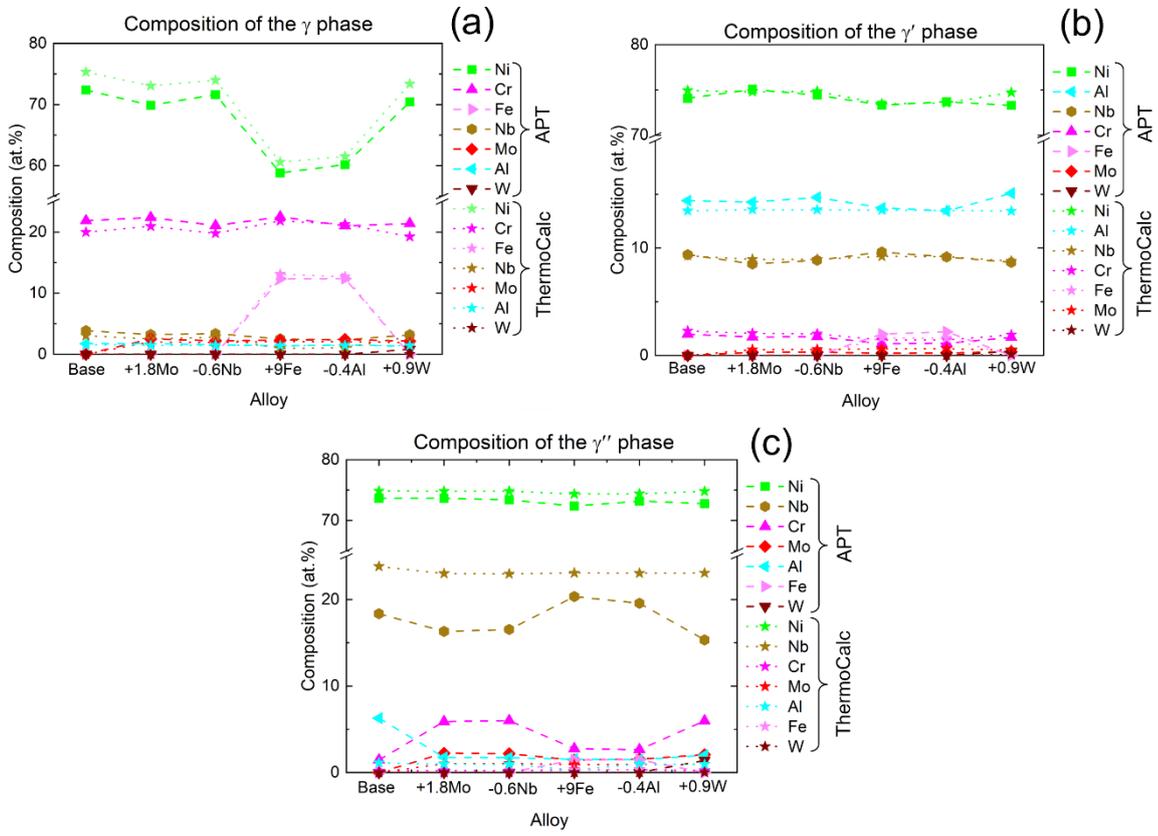

*Figure A1: Phase compositions as measured by APT on samples heat treated at 800 °C vs. Thermo-Calc predictions at 700 °C to compensate for the temperature offset observed in Thermo-Calc predictions.*